# Quantum Squeezing Enhanced Photothermal Microscopy

Pengcheng Fu[1,2]†, Xiao Liu[1]†, Siming Wang[1], Nan Li[2], Chenran Xu[1], Han Cai[2], Huizhu Hu[2], Vladislav V. Yakovlev[3], Xu Liu[2], Shi-Yao Zhu[1,2,4], Xingqi Xu[1]*, Delong Zhang[1]*, and Da-Wei Wang[1,2,4]*

[1]Zhejiang Key Laboratory of Micro-Nano Quantum Chips and Quantum Control, School of Physics, and State Key Laboratory for Extreme Photonics and Instrumentation, Zhejiang University, Hangzhou 310027, Zhejiang Province, China.
[2]College of Optical Science and Engineering, Zhejiang University, Hangzhou 310027, Zhejiang Province, China.
[3]Department of Biomedical Engineering, Texas A&M Univerisy, 77843, College Station, TX, USA
[4]Hefei National Laboratory, Hefei 230088, Anhui Province, China.

*Corresponding authors. Email: xuxingqi@zju.edu.cn, dlzhang@zju.edu.cn, dwwang@zju.edu.cn
†These authors contributed equally: Pengcheng Fu, Xiao Liu.

**Abstract**

Label-free optical microscopy through absorption or scattering spectroscopy provides fundamental insights across biology and materials science, yet its sensitivity remains fundamentally limited by photon shot noise. While recent demonstrations of quantum nonlinear microscopy show sub-shot-limited sensitivity, they are intrinsically limited by availability of high peak-power squeezed light sources. Here, we introduce squeezing-enhanced photothermal (SEPT) microscopy, a quantum imaging technique that leverages twin-beam quantum correlations to detect absorption induced signals with unprecedented sensitivity. SEPT achieves 3.5 dB noise suppression beyond the standard quantum limit, enabling a 2.5-fold increase in imaging throughput or 31% reduction in pump power, while providing an unmatched versatility through the intrinsic compatibility between continuous-wave squeezing and photothermal modulation. We showcase SEPT applications by providing high-precision characterization of nanoparticles and revealing subcellular structures, such as cytochrome c, that remain undetectable under shot-noise-limited imaging. By combining label-free contrast, quantum-enhanced sensitivity, and compatibility with existing microscopy platforms, SEPT establishes a new paradigm for molecular absorption imaging with far-reaching implications in cellular biology, nanoscience, and materials characterization.

**Introduction**

Label-free optical microscopy has transformed our understanding of nature, unveiling the microscopic world through continuous advances in resolution, sensitivity, specificity, and imaging throughput. By harnessing diverse light-matter interactions, including absorption, scattering, and luminescence, it has become an indispensable tool for probing cellular structures[1], molecular metabolic processes[2-5], and dynamic interactions[6,7]. However, detection sensitivity remains fundamentally constrained by the standard quantum limit (SQL), representing the critical bottleneck for high-precision and high-throughput applications. Moreover, excessive illumination induces photodamage and phototoxcitiy to biological specimens, imposing practical constraints on achievable sensitivity.

Squeezed states of light[8], known for its suppressed shot noise, offer a promising approach to surpass the SQL and reduce photodamage in optical measurement. By employing squeezed light as a probe in nonlinear optical imaging modalities, label-free quantum-enhanced stimulated Raman scattering (SRS)[9] and stimulated Brillouin scattering (SBS)[10] has been realized in biological applications, mitigating potential photodamage. However, existing squeezed light sources faces challenges in simultaneously achieving high degree of squeezing and high peak power of the quantum light[9,11], both essential for efficient excitation and detection in nonlinear optical processes. This incompatibility confines quantum nonlinear microscopy to weak-signal regimes at the cost of imaging throughput, making it difficult to surpass state-of-the-art classical SRS[2-5] and SBS[12-14] microscopy, which operate under high-power and low-integration-time conditions.

To unlock the transformative potential of quantum advantages, continuous-wave (cw) squeezed light with high squeezing degrees is ideally suited for linear optical processes, which are independent of the peak power of the light. Photothermal microscopy[15,16], which employs a cw probe beam to detect transient refractive index changes induced by absorption of a modulated pump beam (Fig. 1a), provides an optimal platform for such quantum advantages. Based on linear absorption rather than optical nonlinearities, photothermal microscopy has achieved single nanoparticle and single molecule detection[17,18], as well as metabolic imaging in cells and organisms[19-22]. Despite these achievements, conventional photothermal microscopy remains fundamentally limited by optical shot noise in the probe beam. Efforts to improve sensitivity through extended integration time or increased pump power often compromise signal stability (Supplementary Text, Figure 1,2) and aggravate photodamage, thereby restricting quantitative analysis and real-time monitoring of rapid cellular processes.

To overcome these fundamental limitations, here we present squeezing enhanced photothermal (SEPT) microscopy, which harnesses highly squeezed cw twin beams as quantum-correlated probes while using a classical pump beam. SEPT exploits the inherent compatibility between squeezing bandwidth and optimal photothermal modulation frequency to achieve a 3.5 dB suppression of shot noise without compromising photothermal efficiency. This noise suppression allows 2.5-fold enhancement in imaging throughput, uncovering signatures from weakly absorbing species obscured by shot noise. Moreover, since the pump wavelength can be chosen independently of the quantum probe beams, SEPT maintains broad spectral applicability from visible to mid-infrared wavelengths. We demonstrate its transformative potential through label-free visualization of cytochrome c (Cyt c) in cells and high-precision characterization of nanoparticles. These advances position SEPT as a powerful quantum-enhanced imaging platform for molecular microscopy, with wide-ranging applications in both biology and materials science.

## Results

*Squeezing-enhanced photothermal microscopy*

We developed SEPT microscopy based on the principle illustrated in Figure 1. By exploiting quantum correlation of twin beams to suppress the shot noise, we substantially improved the sensitivity of photothermal microscope (Fig. 1a-c). In conventional coherent photothermal detection, shot noise arises from the discreteness of photon energy and the intrinsic randomness of photon arriving time (Fig. 1b). In quantum-correlated balanced detection, the photons in the twin beams generated simultaneously via a four-wave-mixing (FWM) process exhibit strong temporal intensity correlations. Although the shot noise in either the probe or conjugate beam exceeds that of a coherent field, the noise in their intensity difference is squeezed (Fig. 1c). Importantly, the interaction between the pump and probe lights are mediated by the thermal expansion of the specimen, such that photothermal detection does not rely on the high peak power of ultrafast lasers (Fig. 1a). With a typical modulation frequency below MHz (determined by the μs-level thermal relaxation time), SEPT can fully benefit from the highly squeezed cw twin beams.

We generated quantum-correlated probe and conjugate beams centered at 795 nm through a FWM process in a rubidium ($^{85}$Rb) vapor cell at the $D_1$ transition (see Fig. 1d and Supplementary Text). Using a modified balanced homodyne detector (BHD), we measured up to 7-dB squeezing in the intensity difference of the twin beams (Supplementary Text, Figure 3). To maintain a consistent squeezing degree throughout SEPT imaging, the laser frequency was locked using saturated absorption spectroscopy in a reference rubidium vapor cell (Supplementary Text, Figure 4, 5). The SEPT microscope employs a 532-nm pump laser to induce photothermal effect through electronic absorption. Crutially, the frequency range of the squeezed noise spectra (0.4−4 MHz, Supplementary Figure 3) covered the typical photothermal modulation frequencies[23], enabling both effective noise reduction and efficient photothermal detection. In the experiment, the pump beam was modulated at 719 kHz by an acousto-optic modulator (AOM) to allow for strong quantum correlation and high performance for photothermal imaging.

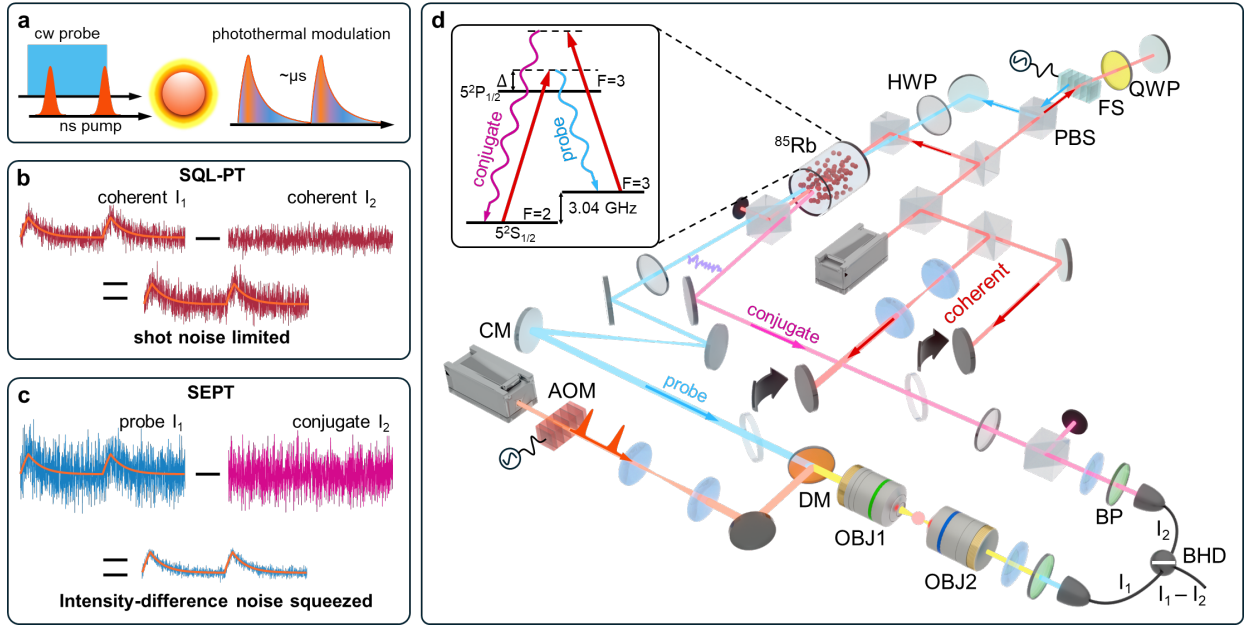

**Fig.1 Principle and implementation of quantum squeezing enhanced photothermal microscope. (a)** Schematic of photothermal (PT) signal generation. Left: a nanosecond pulsed pump beam induces local heating and a cw probe beam detects refractive index modulation. Right: Modulated PT signal. **(b)** Signal and noise in the standard-quantum-limited photothermal detection (SQL-PT) with BHD of coherent fields. **(c)** Signal and noise in SEPT with 7dB two-mode squeezed light. (b) and (c) are numerical simulations with the corresponding photon statistics. **(d)** The experimental setup of SEPT microscope. Quantum-correlated twin beams are generated through FWM process of $^{85}$Rb atoms (inset). To minimize the optical loss, the quantum probe beam is polarization controlled and expanded using a pair of concave mirrors. QWP, quarter-wave plate; HWP, half-wave plate; FS, frequency shifter; PBS, polarizing beamsplitter; CM, concave mirror; AOM, acousto-optic modulator; DM: dichroic mirror; OBJ: objective lens; BP: bandpass filter; BHD: balanced homodyne detector.

A scanning microscope was constructed (Methods) to integrate squeezed-light BHD with photothermal detection while minimizing losses, thereby reducing the leakage of vacuum noise into the measurement. We collinearly combined the photothermal pump and probe beams using a high-efficiency dichroic mirror and focused them through a high-transmission objective. The transmitted probe beam, modulated by the photothermal effect, was collected by a second objective and directed to the BHD after spectral filtering. To realize quantum-correlated photothermal detection, the conjugate beam was independently focused onto the reference photodiode of the BHD, with its intensity controlled by a half-wave plate (HWP) and a polarizing beamsplitter (PBS). A conventional coherent balanced photothermal imaging beam path was inserted through movable mirrors for direct performance comparison. For high-speed image acquisition, we demodulated the BHD output using lock-in detection instead of spectrum analysis, with all imaging parameters, including scanning speed, laser power, and beam routing, controlled through data acquisition (DAQ) hardware. To maintain quantum correlations, we minimized optical loss by precisely controlling the polarization of quantum beams, minimizing the number of optical components, and employing anti-reflection coatings together with index-matched cover glasses (Supplementary Figure 6). These precautionary measures retained 3.5 dB correlation intensity squeezing at 25% total optical loss in the microscope, with the two high-numerical-aperture objectives being the primary source of loss (~10% each).

*Sub-shot-noise detection sensitivity*

To quantitatively assess the quantum advantage in photothermal imaging, we evaluated photothermal signal characteristics by performing sequential spectral and imaging analysis on a single 20-nm gold nanoparticle (AuNP). A clear improvement in imaging quality was observed (Fig. 2a), with a 466-nm diffraction-limited spatial resolution (Supplementary Figure 7), ensuring the sub-cellular imaging performance. Spectral analysis (Fig. 2b, Supplementary Figure 8, Methods) demonstrated that quantum-correlated twin beams reduced the noise floor by 3.5 dB compared to photothermal imaging at SQL (denoted as SQL-PT). The imaging results further indicated that both the average and standard deviations of the background noise in SEPT were consistently lower than those in SQL-PT (Fig. 2c). Importantly, these noise characteristics remained constant when the pump power was adjusted over a broad range, indicating consistent noise suppression in all photothermal measurements. We used the image contrast (Contrast = S / BG, where S is the detected electric signal amplitude, which is proportional to the light intensity, BG is the backgound average amplitude) and signal to backgound ratio (SBR = (S – BG) / $\sigma_{BG}$, where BG is the background average amplitude and $\sigma_{BG}$ is the standard deviation of the background noise) to quantify the imaging performance with a varying pump power. The signal amplitudes in SEPT and SQL-PT had the same linear dependence on the pump power (Supplementary Figure 9), while SEPT achieved a 3.5 dB enhancement in both image contrast (Fig. 2d) and SBR (Fig. 2e) (45% improvement for amplitude ratios defined in our study, corresponding to 110% improvement for power ratios (Contrast$^2$ and SBR$^2$) commonly used in quantum imaging[9,10]). Therefore, SEPT reduces pump power by 31% relative to SQL-PT while maintaining comparable performance, significantly mitigating photodamage and classical noise.

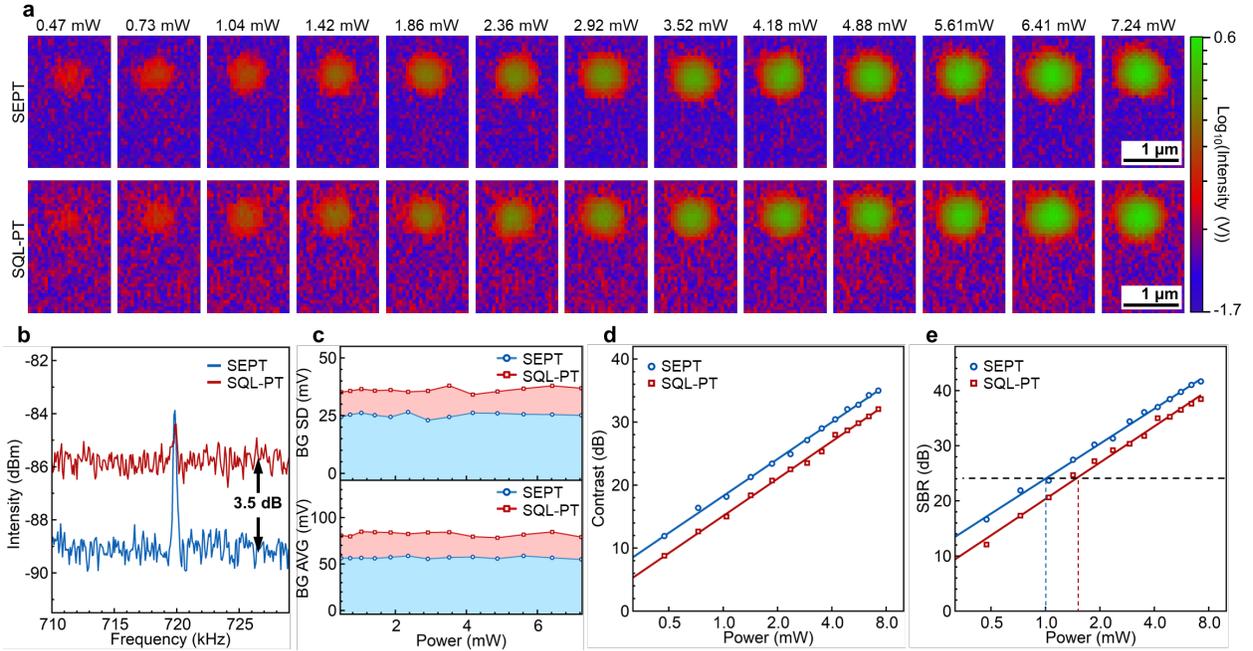

**Fig. 2 Sensitivity improvement in quantum squeezing enhanced photothermal microscopy.** (**a**) Comparison of SEPT and SQL-PT images of a 20-nm AuNP under various pump power. Scale bars: 1 μm. (**b**) Spectral analysis of the photothermal signal measured using quantum-correlated twin beams (blue line) and a coherent beam (red line). Spectra were averaged over 10 datasets, each for 5 seconds integration time, 300-Hz resolution bandwidth, and 30-Hz video bandwidth. (**c**) Background (BG) noise characteristics derived from panel a, showing the mean (AVG) and standard deviations (SD). (**d-e**) Comparison between image comtrast (**d**) and SBR (**e**) of SEPT and SQL-PT across varying pump power. The decibel scales of Contrast and SBR were obtained by $10\log_{10} x^2$ (dB), where $x$ is the correspoinding amplitude ratio.

To further evaluate the quantum advantage of SEPT, we systematically characterize signal quality metrics as a function of integration time and pump power. Both photothermal intensity and phase signals were simultaneously recorded after lock-in demodulation (Methods). Contour plots (Fig. 3a-b) and line profiles (Fig. 3c) consistently showed that SEPT achieved a faster reduction in phase uncertainty than the SQL-PT detection until a long integration time (10 s), underscoring its ability in precise measurement of thermal diffusivity[24,25]. Across all tested conditions, SEPT exhibited a constant improvement in Contrast (Supplementary Figure 10) and SBR, demonstrating robust signal extraction capability under various noise conditions (Fig. 3d). In contrast, the SNR = (S − BG) / $\sigma_S$, where $\sigma_S$ is the standard deviation of signal amplitude during the integration time, plateaued at long integration time and large pump power, due to increased signal fluctuations (Fig. 3e). This signal fluctuation arises from classical noises, including low-frequency power drifting in the pump beam and hot Brownian motion of heated nanoparticles[26,27], which cannot be suppressed by quantum enhancement. These classical noise sources limit both image quality and throughput, underscoring the need to improve imaging performance without raising the illumination power. SEPT was able to achieve signal quality comparable to that of SQL-PT while boosting imaging throughput by a factor of 2.5 (Fig. 3f), thereby effectively reducing the integration time and circumventing the classical noise while minimizing photodamage. These capabilities demonstrated SEPT's potential for high-precision, high-temporal-resolution studies of dynamic biological processes with reduced perturbation to living samples.

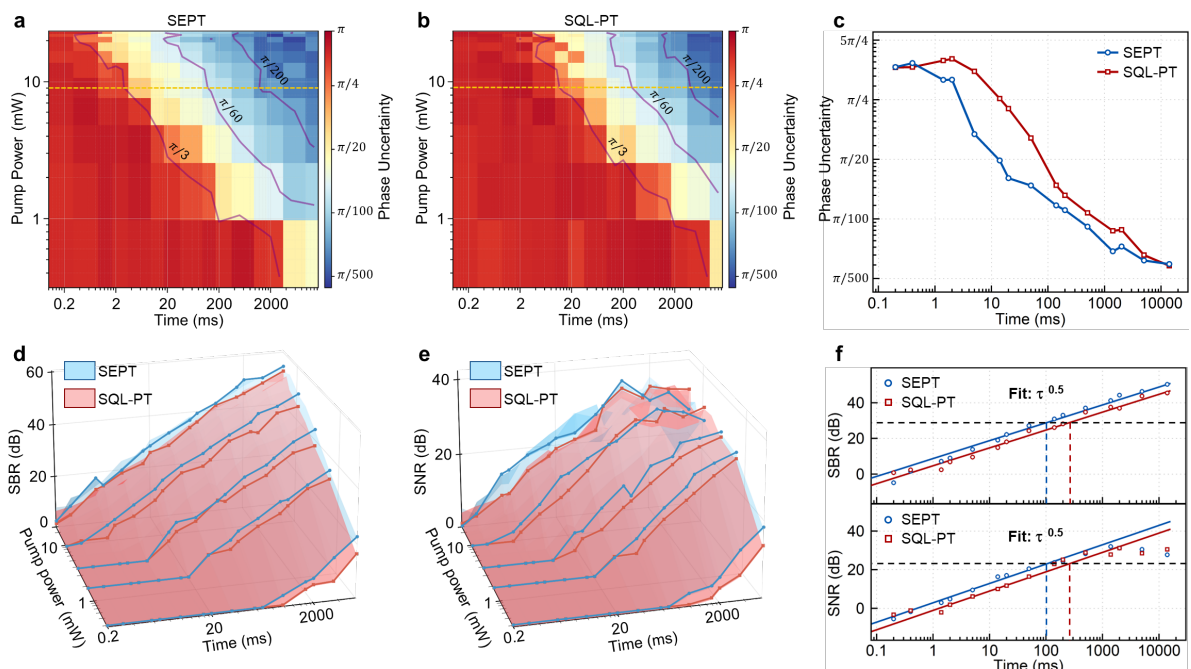

**Fig. 3 Phase stability and signal quality enhancement in squeezing enhanced photothermal microscopy.** (**a-b**) Photothermal phase uncertainty measured by SEPT (a) and SQL-PT (b) microscopy. Phase signals were obtained from the phase channel of the lock-in amplifier. (**c**) Phase uncertainty profiles along the yellow-dashed lines in (a) and (b). (**d-e**) Comparison of the SBR (d, noise from background) and SNR (e, noise from signal) between SEPT and SQL-PT. In (d-e**)**, the data are directly connected by lines to guide the eyes. (**f**) The SBR and SNR in SQL-PT (red squares) and SEPT (blue circles) as functions of integration time at 9.1-mW pump power. The solid lines in (f**)** are the fitting results of power-law function with a fixed exponent of 0.5. The dashed lines in (f) indicate the integration time required to achieve equivalent SBR and SNR levels.

*High-precision size analysis of nanoparticles*

With the quantum enhanced SBR and SNR, we demonstrated SEPT's advantages of quantitative analysis by discrimination of AuNPs with two different sizes that were indistinguishable by coherent light (Fig. 4). We compared the two techniques in imaging a 1:1 mixture of 13-nm and 15-nm diameter AuNPs. Quantum-correlated detection provided substantial noise suppression in both photothermal intensity and phase channels (Fig. 4a-d), revealing multiple weak signals of AuNPs buried in the shot noise of laser (Figs. 4e-f, zoomed-in views and line profiles between the two white arrows).

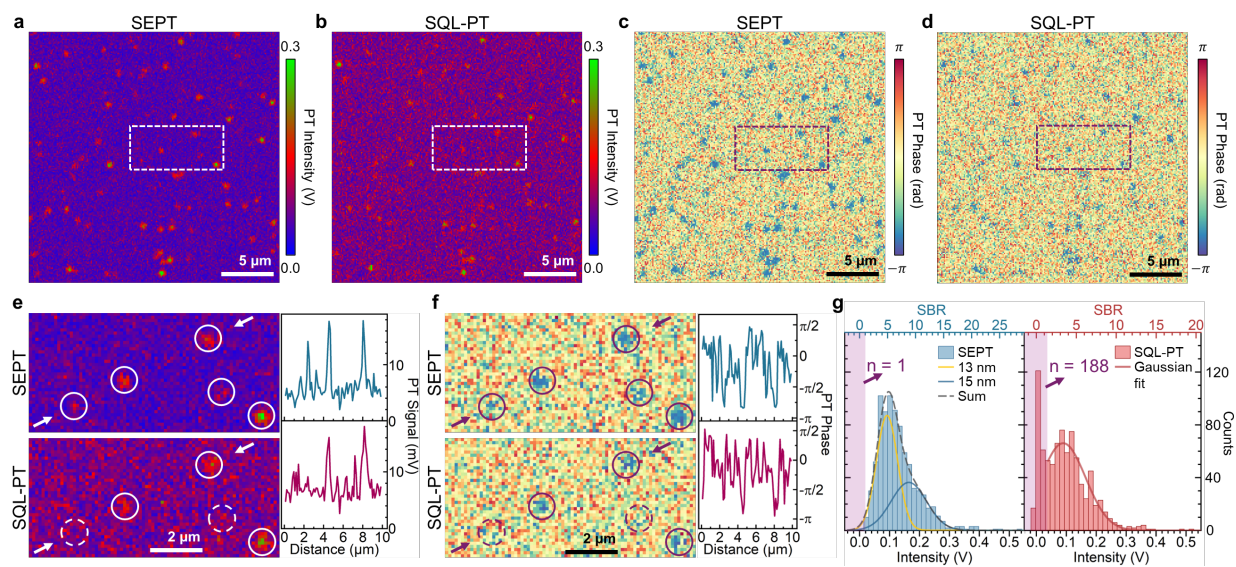

**Fig. 4 Quantum-enhanced nanoparticle characterization.** (**a-b**) Photothermal amplitude images of AuNPs obtained by SEPT (a) and SQL-PT (b). (**c-d**) Photothermal phase images acquired with SEPT (c) and SQL-PT (d). Scale bars: 5 μm. (**e**) Magnified views of SEPT (top panel, from dashed box in a) and SQL-PT (bottom panel, from dashed box in b) with intensity profiles along arrow pairs (right panels). (**f**) Magnified phase images from dashed-box regions in panels (c) and (d) with corresponding phase profiles. Scale bars: 2 μm. (**g**) Photothermal intensity distributions extracted from SEPT and SQL-PT measurements. Particle sizes determined via multi-peak Gaussian fitting. A total of 836 particles were analyzed.

Statistical analysis demonstrated SEPT's quantitative advantages (Fig. 4g). We analyzed the intensity and SBR of 836 nanoparticles using both SEPT and SQL-PT. SEPT detected 187 additional particles that were submerged in the noise of SQL-PT (purple-shaded region, defined by SBR < 1). Critically, SEPT's reduced noise floor enabled discrimination between 13-nm and 15-nm particle populations. The SEPT-derived intensity histogram exhibited superior smoothness and revealed a distinct shoulder peak delineating the two populations. Multi-peak Gaussian fitting yielded an intensity ratio of 1.80 between the two populations, corresponding to a diameter ratio of 15.6:12.8 nm based on the cubic scaling relationship[23] between photothermal signal and particle diameter (Supplementary Figure 11). This measurement agreed with the actual mean diameters (15.6 ± 1 nm and 12.7 ± 0.8 nm, respectively) determined by transmission electron microscopy (Supplementary Figure 12). Furthermore, achieving comparable SNR in SQL-PT necessitates higher pump powers that may induce thermal instabilities, thereby compromising measurement precision. These results demonstrate that SEPT enables superior size discrimination and quantitative characterization of nanoparticle heterogeneity.

*Quantum-enhanced biological imaging*

To demonstrate SEPT's capability for biomolecular imaging, we performed label-free visualization of Cyt c in mammalian cells. Cyt c, a heme-containing mitochondrial protein, regulates critical cellular processes including respiration and apoptosis[28,29]. Current approaches for studying *in situ* mitochondria and Cyt c primarily rely on immunofluorescence, which can disrupt native morphology[30], impair protein functions[31], and generate artifacts in mitochondrial transfer studies[32]. While label-free Raman microscopy has been demonstrated for visualizing Cyt c[33], its weak scattering cross-section limits both sensitivity and imaging speed while causing phototoxicity for extended observation time. SEPT overcomes these limitations through quantum-enhanced detection of Cyt c's absorption.

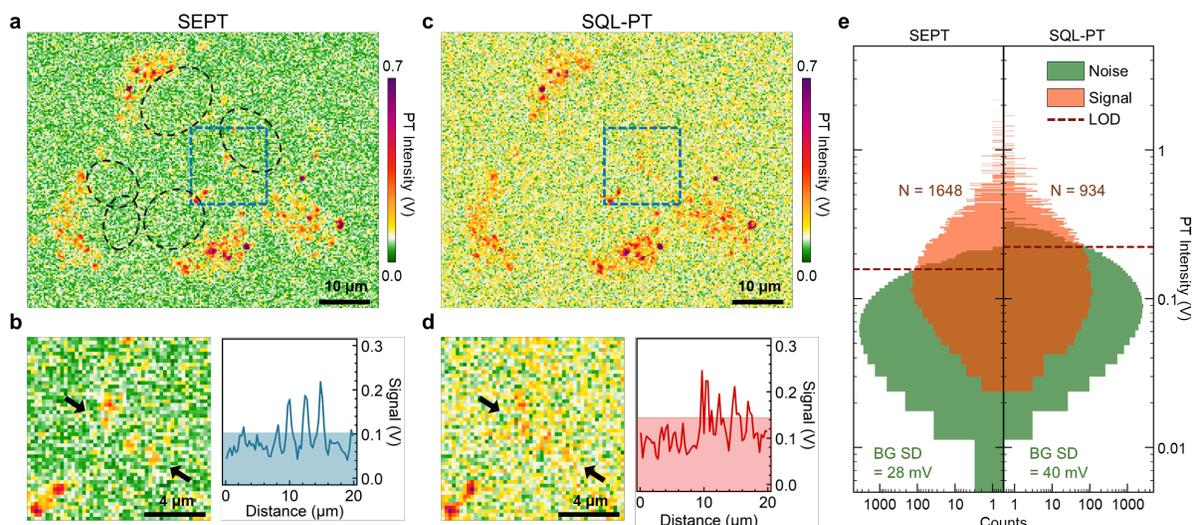

**Fig. 5. Quantum enhanced imaging of cytochrome c in HEK293 cell.** (**a**) SEPT imaging revealing subcellular Cyt c distribution. Scale bar: 10 μm. (**b**) Magnified view of blue-dashed region in (A) with intensity profile (right panel) along the arrow pairs. Scale bars: 4 μm. (**c-d**) SQL-PT results in the same areas as in (a) and (b). The blue and red areas in the line profiles indicate an SBR below 1. (**E**) Statistical distribution of signal (orange) and noise (green) levels comparing SEPT and SQL-PT performance. The limit of detection (LOD) is defined by SBR = 3 (LOD = BG + $3\sigma_{BG}$).

We imaged Cyt c in fixed HEK293 cells (Methods) by exploiting its absorption band at 532 nm[34]. In Fig. 5, we compared SEPT with SQL-PT under identical conditions. SEPT revealed substantially more structural details, with enhanced contrast enabling visualization of subcellular features obscured by the shot noise in SQL-PT imaging (Figs. 5a-d). Nuclear regions, identified through co-registration with brightfield imaging (Supplementary Figure 13, 14), appeared as expected voids in the Cyt c distribution (black-dashed circles, Fig. 5a).

Furthermore, quantitative analysis confirmed SEPT's superior performance in extracting signal from background noise (Fig. 5e, the extraction mask shown in Supplementary Figure 15). While both techniques yielded comparable signal intensity distributions (orange histograms), SEPT exhibited a narrower background noise distribution with a lower standard deviation (28 mV) compared to 40 mV of SQL-PT (green histograms). The limit of detection (LOD), defined as LOD = BG + $3\sigma_{BG}$ (corresponding to SBR = 3), was used to quantify the number of effective pixels. This 30% reduction in LOD (158 mV for SEPT vs. 223 mV for SQL-PT) enabled SEPT to resolve 77% more spatial features (1,648 vs. 934 pixels exceeding their respective LOD thresholds), revealing cellular structures previously obscured by shot noise. These findings demonstrated that SEPT achieved the sensitivity required for molecular imaging while maintaining the spatial resolution and acquisition speed essential for biological applications.

**Discussion**

The integration of quantum-correlated twin beams with photothermal detection marks a clear advance in quantum-enhanced microscopy. Unlike nonlinear approaches that demand both intense pulsed excitation and highly squeezed light, SEPT leverages linear absorption processes to fully harness the high squeezing degree of cw twin beams from FWM. The 3.5 dB noise suppression achieved by SEPT translates to either a 31% reduction in required pump power or 2.5-fold imaging speed increase, which addresses a fundamental challenge in biological microscopy, the trade-off between sensitivity and photodamage. This quantum enhancement is particularly significant because it reveals previously undetectable weak absorbers and enables discrimination of subtle molecular heterogeneities, suggesting applications in many biological processes obscured by shot noise.

The separation of classical pump, which provides exceptional chemical selectivity, from quantum probe, which provides a disruptive improvement in detection sensitivity, represents a key conceptual advancement in the presented quantum imaging. While commercial laser sources are available across a broad frequency range, highly squeezed light sources remain restricted to specific wavelengths. The novel photothermal detection strategy enables flexible spectral selectivity for absorption measurements across the entire optical spectrum while preserving high levels of noise squeezing. Therefore, SEPT circumvents the need for wavelength-tunable quantum light sources for correlated detection[35,36]. By varying the pump wavelength from visible to mid-infrared, SEPT accesses diverse molecular signatures ranging from electronic transitions[18] to the fundamental and overtone absorption of molecular vibrations[19,37], which is crucial for multimodal imaging applications where diverse molecular signatures must be probed simultaneously.

SEPT establishes a versatile framework whereby highly squeezed cw light resources can be readily deployed into diverse photothermal-based techniques, overcoming fundamental sensitivity barriers in interdisciplinary research. The demonstrated phase-stabilized detection capabilitiy (Fig. 3a) allows higher precision in phase sensitive applications, such as enantiomer discrimination through photothermal detection of circular dichroism[38]. By detecting photothermal effect from nonlinear optical processes such as two-photon[39] and stimulated Raman[40] excitations, SEPT can be extended to nonlinear optical microscopy, thereby offering broader spectral information than linear absorption. Our approach is ready to be extended to detect the photothermal effect of nonlinear absorption by using the same strategy of classical pump and quantum probe. The performance of SEPT can be further improved by enhancing detection efficiency and employing twin beams with higher squeezing, which was realized in phase-sensitive amplifier through interference[41]. Reducing optical losses, particularly in objectives, is critical for preserving squeezing and maximizing quantum enhancement. Without using objectives we can adopt cantilever-based photothermal detection[42], as experimentally demonstrated in quantum-enhanced beam displacement sensing with squeezed light[43,44].

**Methods**

**SEPT microscope.** The imaging system was constructed on a microscope frame (BX51, Olympus) with the following components. The cw pump beam (Lighthouse Photonics, Sprout) was modulated with an AOM (Gooch & Housego, 3080-125) for PT excitation. To minimize the loss of squeezed light caused by multiple reflection elements, the polarization of the probe beam was rotated to s state with a HWP. The probe beam was expanded with two concave mirrors, and colinearly combined with the pump beam with a dichroic mirror (Thorlabs, DMLP650) and focused with a high-transmission objective (Leica, HC PLAPO 20x/0.80, NA = 0.8). Then the probe beam was forward-detected by a modified BHD (Thorlabs, PDB440A, with the photodiode detector from Hamamatsu Photonics S3399) after being collected by the other objective (Leica, HC PLAPO 63x/1.40-0.60 OIL, NA = 1.4). A bandpass filter (Thorlabs, FBH800-40) was employed to remove the pump light and suppress ambient noise. To measue the quantum enhancement, a coherent probe beam was inserted by a mirror mounted on a motorized stage (Zolix, LXC50-80100). Samples were sandwiched with coverslips with single-side coated (Supplementary Figure 6). Raster scanning of samples (Physik Instrumente, P-562.3) and signal acquisition were controlled by a DAQ card (National Instruments, 6363) with a custom-built code. The PT signals were analyzed with a spectrum analyzer or demodulated with a lock-in amplifier (Zurich instrument, HF2LI).

**Spectrum acquisition.** The photovoltage signal was sent to a spectrum analyzer (Rigol, DSA875) and the intensity noise in the frequency range of 0–4 MHz was acquired through cluster communication port with a computer.

**Single particle PT-signal assessment.** For photovoltage signal acquisition from single particles, the sampling focus was aligned with the center of the gold nanoparticles (AuNPs). The resulting photovoltage signal was then directed to a lock-in amplifier for demodulation. A custom-written code was employed to acquire the photothermal signal while systematically varying the lock-in amplifier's time constant and increasing the heating power. The latter was achieved by rotating a HWP (Jcoptix, MWP22H-532M) with a motorized rotator (Thorlabs, K10CR2/M).

**AuNPs imaging.** AuNPs were diluted in a 0.4% polyvinyl alcohol (PVA) solution with a volume ratio of 1:10, then a 100 μL of the suspension was spin-coated onto a coverslip at 500 rpm for 30 s and 2000 rpm for 60 s. The sample was immersed with 30 μL silicone oil (see Supplementary Figure 11a) and sealed with a single-side coated cover glass for imaging (see Supplementary Figure 6a).

**Statistical analysis of AuNP clusters signal.** The masks that were used to identify AuNP positions were generated from the high-SBR SEPT images (see Supplementary Figure 14). These masks were subsequently applied to both SEPT and SQL-PT image sets to extract signal intensities from the corresponding regions. A total of 836 AuNPs were analyzed across all imaging sets. The SBR values for the AuNPs in each individual image were calculated separately based on the local background mean and standard deviation derived from the same image.

**Cell culture and imaging.** HEK 293T cells were cultured on the 170-μm cover glass in Dulbecco Modified Eagle Medium (DMEM, ThermoFisher) supplemented with 10% fetal bovine serum. Prior to imaging, the cell was fixed with 4% formalin for 15 min, followed by washing with phosphate-buffered saline. The samples were then sealed by a single-side coated cover glass and immersed in phosphate-buffered saline for imaging.

**Acknowledgments:** We are grateful to Jinze Wu and Yongqing Zhang for upgrading the BHDs and preparing the materials. We also thank Zhongzhong Qin, Shengshuan Liu and Jietai Jing for their suggestions in improving our squeezed light source. This work was supported by the National Key Research and Development Program of China (2024YFA1408900), the National Natural Science Foundation of China (12404573, 12434020, 12325412, 12374335, U21A20437, 11934011), the Fundamental Research Funds for the Central Universities (2024FZZX02-01-02),

## References

1. Sigal, Y. M., Zhou, R. & Zhuang, X. Visualizing and discovering cellular structures with super-resolution microscopy. *Science* **361**, 880-887 (2018).
2. Freudiger, C. W. *et al.* Label-free biomedical imaging with high sensitivity by stimulated Raman scattering microscopy. *Science* **322**, 1857-1861 (2008).
3. Cheng, J.-X. & Xie, X. S. Vibrational spectroscopic imaging of living systems: An emerging platform for biology and medicine. *Science* **350**, aaa8870 (2015).
4. Hu, F., Shi, L. & Min, W. Biological imaging of chemical bonds by stimulated Raman scattering microscopy. *Nat. Methods* **16**, 830-842 (2019).
5. Min, W., Cheng, J.-X. & Ozeki, Y. Theory, innovations and applications of stimulated Raman scattering microscopy. *Nat. Photonics*, 1-14 (2025).
6. Park, Y., Depeursinge, C. & Popescu, G. Quantitative phase imaging in biomedicine. *Nat. Photonics* **12**, 578-589 (2018).
7. Ginsberg, N. S., Hsieh, C.-L., Kukura, P., Piliarik, M. & Sandoghdar, V. Interferometric scattering microscopy. *Nat. Rev. Methods Primers* **5**, 23 (2025).
8. Walls, D. F. Squeezed states of light. *Nature* **306**, 141-146 (1983).
9. Casacio, C. A. *et al.* Quantum-enhanced nonlinear microscopy. *Nature* **594**, 201-206 (2021).
10. Li, T., Cheburkanov, V., Yakovlev, V. V., Agarwal, G. S. & Scully, M. O. Harnessing quantum light for microscopic biomechanical imaging of cells and tissues. *Proc. Natl. Acad. Sci. U. S. A.* **121**, e2413938121 (2024).
11. Xu, Z., Nitanai, S., Oguchi, K. & Ozeki, Y. Pushing the sensitivity of stimulated Raman scattering microscopy with quantum light: Current status and future challenges. *Appl. Phys. Lett.* **127**, 040501 (2025).

12   Yang, F. *et al.* Pulsed stimulated Brillouin microscopy enables high-sensitivity mechanical imaging of live and fragile biological specimens. *Nat. Methods* **20**, 1971-1979 (2023).
13   Shaashoua, R. *et al.* Brillouin gain microscopy. *Nat. Photonics* **18**, 836-841 (2024).
14   Qi, Y. *et al.* Stimulated Brillouin scattering microscopy with a high-peak-power 780-nm pulsed laser system. *Nat. Photonics*, 1-9 (2025).
15   Adhikari, S. *et al.* Photothermal microscopy: imaging the optical absorption of single nanoparticles and single molecules. *ACS Nano* **14**, 16414-16445 (2020).
16   Bai, Y., Yin, J. & Cheng, J.-X. Bond-selective imaging by optically sensing the mid-infrared photothermal effect. *Sci. Adv.* **7**, eabg1559 (2021).
17   Lasne, D. *et al.* Single nanoparticle photothermal tracking (SNaPT) of 5-nm gold beads in live cells. *Biophys. J.* **91**, 4598-4604 (2006).
18   Gaiduk, A., Yorulmaz, M., Ruijgrok, P. & Orrit, M. Room-temperature detection of a single molecule's absorption by photothermal contrast. *Science* **330**, 353-356 (2010).
19   Zhang, D. *et al.* Depth-resolved mid-infrared photothermal imaging of living cells and organisms with submicrometer spatial resolution. *Sci. Adv.* **2**, e1600521 (2016).
20   Fu, P. *et al.* Super-resolution imaging of non-fluorescent molecules by photothermal relaxation localization microscopy. *Nat. Photonics* **17**, 330-337 (2023).
21   He, H. *et al.* Mapping enzyme activity in living systems by real-time mid-infrared photothermal imaging of nitrile chameleons. *Nat. Methods* **21**, 342-352 (2024).
22   Fu, P. *et al.* INSPIRE: Single-beam probed complementary vibrational bioimaging. *Sci. Adv.* **10**, eadm7687 (2024).
23   Berciaud, S., Cognet, L., Blab, G. A. & Lounis, B. Photothermal Heterodyne Imaging of Individual Nonfluorescent Nanoclusters and Nanocrystals. *Phys. Rev. Lett.* **93**, 257402 (2004).
24   Heber, A., Selmke, M. & Cichos, F. Thermal diffusivities studied by single-particle photothermal deflection microscopy. *ACS Photonics* **4**, 681-687 (2017).
25   Samolis, P. D. & Sander, M. Y. Phase-sensitive lock-in detection for high-contrast mid-infrared photothermal imaging with sub-diffraction limited resolution. *Opt. Express* **27**, 2643-2655 (2019).
26   Radunz, R., Rings, D., Kroy, K. & Cichos, F. Hot Brownian particles and photothermal correlation spectroscopy. *J. Phys. Chem. A* **113**, 1674-1677 (2009).
27   Rings, D., Schachoff, R., Selmke, M., Cichos, F. & Kroy, K. Hot brownian motion. *Phys. Rev. Lett.* **105**, 090604 (2010).
28   Ow, Y.-L. P., Green, D. R., Hao, Z. & Mak, T. W. Cytochrome c: functions beyond respiration. *Nat. Rev. Mol. Cell Biol.* **9**, 532-542 (2008).
29   Hüttemann, M. *et al.* The multiple functions of cytochrome c and their regulation in life and death decisions of the mammalian cell: From respiration to apoptosis. *Mitochondrion* **11**, 369-381 (2011).
30   Lee, C.-H., Wallace, D. C. & Burke, P. J. Photobleaching and phototoxicity of mitochondria in live cell fluorescent super-resolution microscopy. *Mitochondrial Commun.* **2**, 38-47 (2024).
31   Alford, R. *et al.* Toxicity of organic fluorophores used in molecular imaging: literature review. *Mol. Imag.* **8**, 7290.2009. 00031 (2009).
32   Brestoff, J. R. *et al.* Recommendations for mitochondria transfer and transplantation nomenclature and characterization. *Nat. Metab.* **7**, 53-67 (2025).
33   Okada, M. *et al.* Label-free Raman observation of cytochrome c dynamics during apoptosis. *Proc. Natl. Acad. Sci. U. S. A.* **109**, 28-32 (2012).

| | |
|---|---|
| 34 | Wang, L., Santos, E., Schenk, D. & Rabago-Smith, M. Kinetics and mechanistic studies on the reaction between cytochrome c and tea catechins. *Antioxidants* **3**, 559-568 (2014). |
| 35 | Brida, G., Genovese, M. & Ruo Berchera, I. Experimental realization of sub-shot-noise quantum imaging. *Nat. Photonics* **4**, 227-230 (2010). |
| 36 | Moreau, P.-A., Toninelli, E., Gregory, T. & Padgett, M. J. Imaging with quantum states of light. *Nat. Rev. Phys.* **1**, 367-380 (2019). |
| 37 | Ni, H. *et al.* Millimetre-deep micrometre-resolution vibrational imaging by shortwave infrared photothermal microscopy. *Nat. Photonics* **18**, 944-951 (2024). |
| 38 | Spaeth, P. *et al.* Circular dichroism measurement of single metal nanoparticles using photothermal imaging. *Nano Lett.* **19**, 8934-8940 (2019). |
| 39 | Lu, S., Min, W., Chong, S., Holtom, G. R. & Xie, X. S. Label-free imaging of heme proteins with two-photon excited photothermal lens microscopy. *Appl. Phys. Lett.* **96**, 113701 (2010). |
| 40 | Zhu, Y. *et al.* Stimulated Raman photothermal microscopy toward ultrasensitive chemical imaging. *Sci. Adv.* **9**, eadi2181 (2023). |
| 41 | Liu, S., Lou, Y. & Jing, J. Interference-induced quantum squeezing enhancement in a two-beam phase-sensitive amplifier. *Phys. Rev. Lett.* **123**, 113602 (2019). |
| 42 | Dazzi, A. & Prater, C. B. AFM-IR: Technology and applications in nanoscale infrared spectroscopy and chemical imaging. *Chem. Rev.* **117**, 5146-5173 (2017). |
| 43 | Treps, N. *et al.* A quantum laser pointer. *Science* **301**, 940-943 (2003). |
| 44 | Pooser, R. C. & Lawrie, B. Ultrasensitive measurement of microcantilever displacement below the shot-noise limit. *Optica* **2**, 393-399 (2015). |

# Supplementary Materials for
## Quantum Squeezing Enhanced Photothermal Microscopy

Pengcheng Fu[1,2]†, Xiao Liu[1]†, Siming Wang[1], Nan Li[2], Chenran Xu[1], Han Cai[2], Huizhu Hu[2], Vladislav V. Yakovlev[3], Xu Liu[2], Shi-Yao Zhu[1,2,4], Xingqi Xu[1]*, Delong Zhang[1]*, and Da-Wei Wang[1,2,4]*

Corresponding authors: xuxingqi@zju.edu.cn, dlzhang@zju.edu.cn, dwwang@zju.edu.cn

**The PDF file includes:**

> Materials and Methods
> Supplementary Text
> Figs. S1 to S14

**Supplementary Text**

Signal characterization of SQL-PT.

We evaluated photothermal signal characteristics from AuNPs by varying integration time and heating power, as discussed above in Materials and Methods. While both SNR (defined as SNR = (S – BG) / $\sigma_S$) and SBR (defined as SBR = (S – BG) / $\sigma_{BG}$) were expected to be proportional to the heating power and the square root of integration time[1], practical experiments face multiple constraints which led to deviations from theoretical dependencies at overdosed-heating conditions. Specifically, the background noise was elevated due to the fluctuations of medium and/or substrates, while the PT signals also became unstable under high heating power. The former compromised the SBR and the latter severely degraded the SNR, as shown in Figs. S1 and S2.

Consistent with previous reports[2-4], such instability were primarily attributed to hot Brownian motion[5,6] and nonlinear PT response induced by overheating[7,8], which posed substantial challenges for both image quality and acquisition throughput. It called for improving imaging performance under conventional illumination conditions.

Intensity-difference squeezed light generation in atomic ensemble.

The light beam from a narrow-linewidth laser (Toptica, DLC TA PRO 795) was split to two beams (the pump beam and the probe beam) with a polarizing beamsplitter. The horizontally polarized beam – served as the probe – was red-shifted by 3.036 GHz through double passing the FS (Brimrose Corp, TEF-1500-100-795). The two beams were filtered with single-mode fibers and then weakly focused into a 12.5 mm-long $^{85}$Rb cell with a 0.3° angle intersect. The powers of the pump and the probe for FWM process were 300 mW and 0.1 mW, and the beam sizes at the focal point were approximately 700 μm and 400 μm, respectively. The cell was heated to 112±0.1 °C and the laser frequency is detuned from the $^{85}$Rb $D_1$ line ($5S_{1/2}$ (F=2) → $5P_{1/2}$ (F=3)) by $\Delta \approx 0.8$ GHz. The atoms were driven in the close-loop transition of the FWM (Supplementary Figure 1d), which generated the two-mode squeezed light. The residual pump was filtered by a polarizing beamsplitter and multiple irises.

Characterization of squeezed light source.

The AC output from the customized BHD was fed directly into a spectral analyzer for noise measurements of both classical and squeezed light. The directly measured squeezing level was 6.4 dB (Supplementary Figure 3a). To eliminate the influence of residual coherent pump and electronic noise, we evaluated the actual squeezing degree with the power dependence measurements[9,10]. This analysis yielded a squeezing level of 7 dB (Supplementary Figure 3b).

Frequency locking with frequency shifted saturation absorption of $^{85}$Rb atom.

Both the laser frequency and the power drift could notably deteriorate the squeezing. To maintain high stability of the squeezed light, we employed an internal modulation scheme based on a frequency-shifted saturation absorption spectrum (SAS) to lock the laser frequency (Supplementary Figure 4). Since the FWM process required a single-photon detuning ($\Delta$), we introduced another AOM (CETC, SGT400-795-0.1TA-B100) to red-detune the laser frequency to be in resonance with the $^{85}$Rb $D_1$ line ($5S_{1/2}$ (F=2) → $5P_{1/2}$ (F=3)). This configuration allowed the laser frequency to be locked to the atomic transition while maintaining a desired detuning. To characterize the detuned frequency, we compared the frequency-shifted SAS with the original reference spectrum (Supplementary Figure 4b). As shown in Supplementary Figure 4c, the free-running laser frequency exhibited drifts exceeding 100 MHz, whereas in the locked condition, the frequency fluctuation remained within 2 MHz.

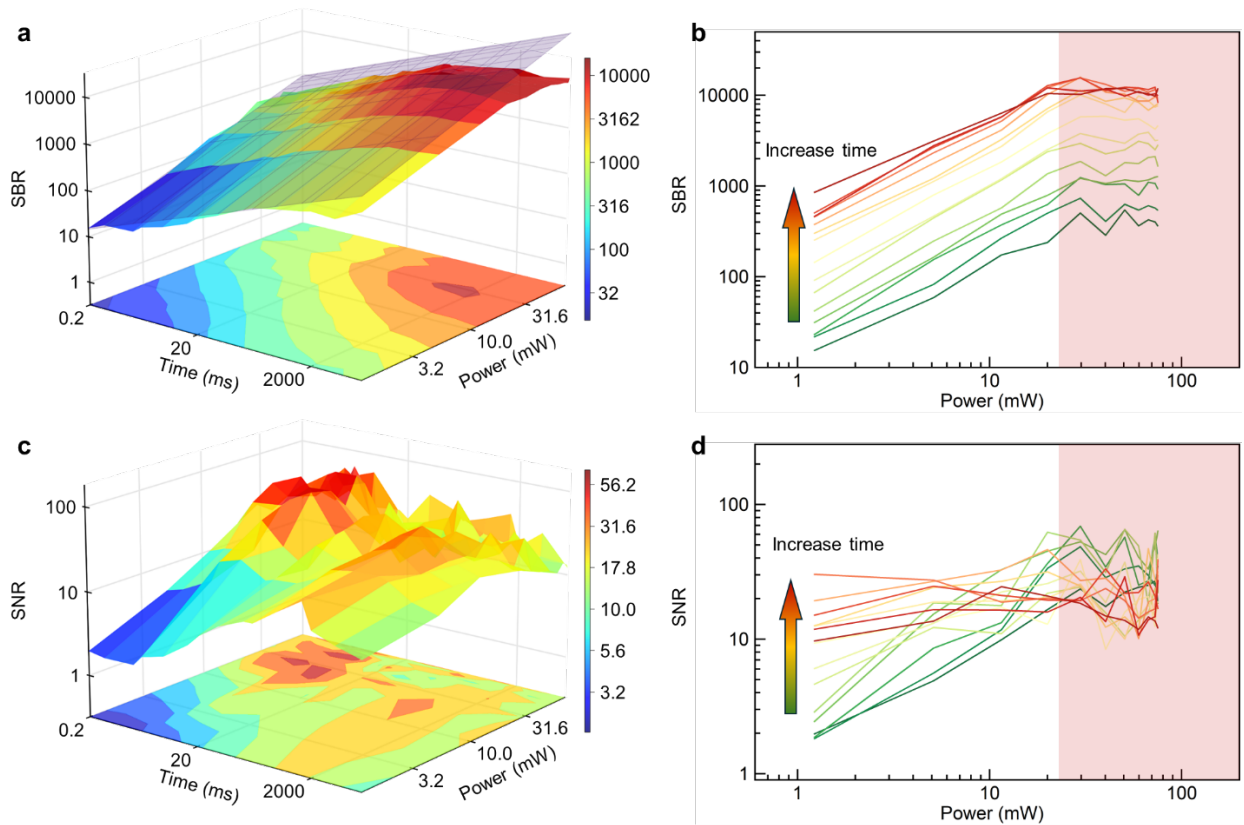

**Fig. S1. Characterization of the PT signals under varying heating power and integration time.** (A) and (B) show the signal-to-background ratio (SBR), revealing a saturation region at high power. (C) and (D) illustrate the signal-to-noise ratio (SNR), which is affected by signal fluctuations under high power and prolonged integration time.

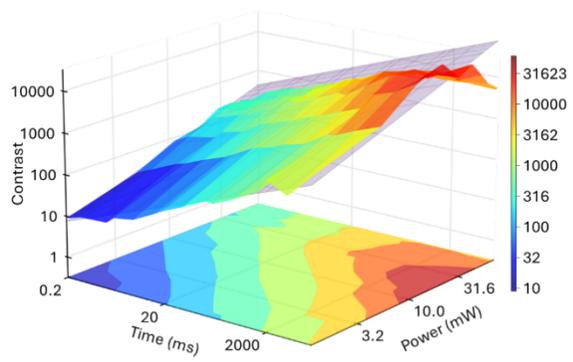

**Fig. S2.** PT signal contrast at different integration time and heating power.

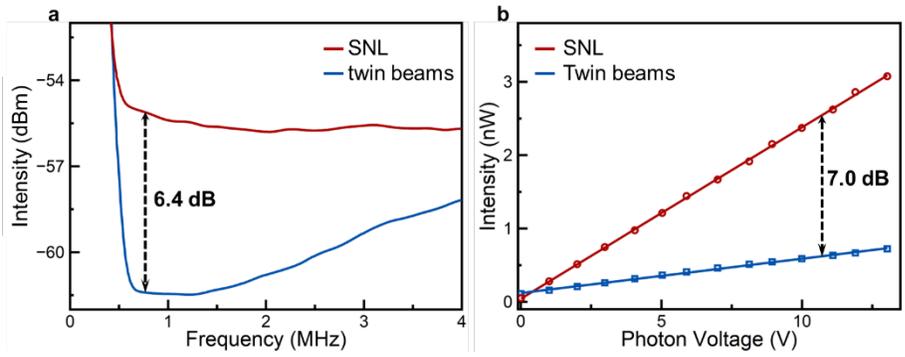

**Fig. S3. Characterization of the squeezed light source.** (A) Comparison of the spectrum noise between the coherent field (red) and the two-mode squeezed laser (blue). (B) Laser noise power as a function of optical power.

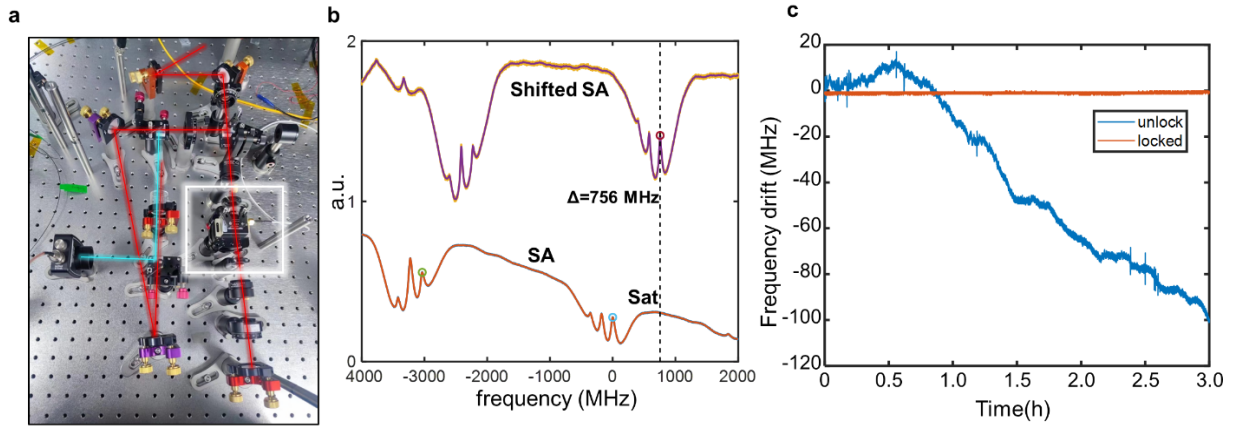

**Fig. S4. Frequency locking system.** (A) Schematic of the optical path for laser frequency stabilization using the saturated absorption spectrum (SAS) of $^{85}$Rb. An AOM is used to generate a detuned beam for locking. (B) The resulting SAS with a frequency offset of approximately 760 MHz. (C) Comparison of laser frequency drift in 3 hours between unlocked (free-running) and locked modes.

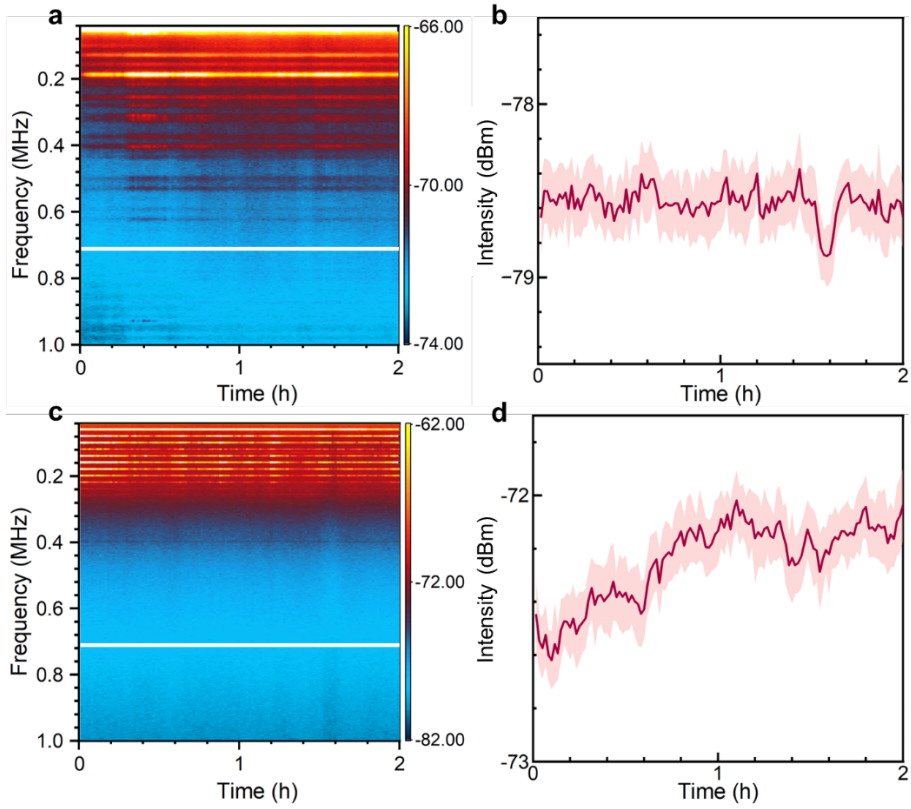

**Fig. S5. Highly robust squeezed light enabled by frequency locking.** (A) Spectrum noise before frequency locking in 2 hours. (B) Intensity noise at 0.70 - 0.72 MHz in 2 hours, indicated by the white line in (A). (C) Stabilized spectrum noise after frequency locking in 2 hours. (D) Intensity noise at 0.70 - 0.72 MHz, corresponding to the white line in (C).

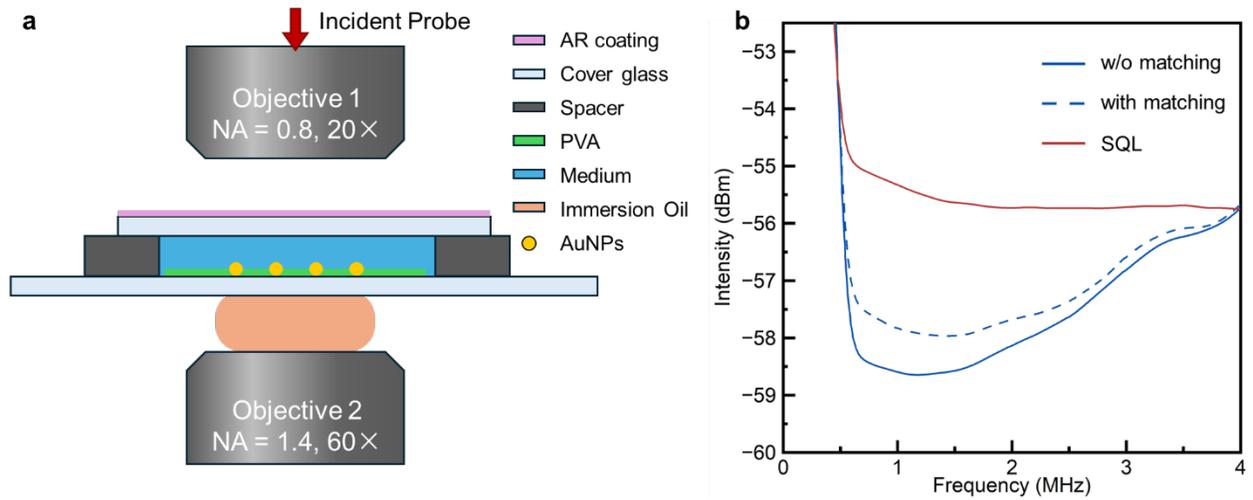

**Fig. S6. Improved transmission of the SEPT quantum probe.** (A) index-matching and anti-reflection coating configuration. (B) Comparison of intensity noise of twin beams before and after the transmission optimization. AR, Anti-reflection, PVA: polyvinyl alcohol. With this optimized scheme, losses from reflection of glasses is negligible.

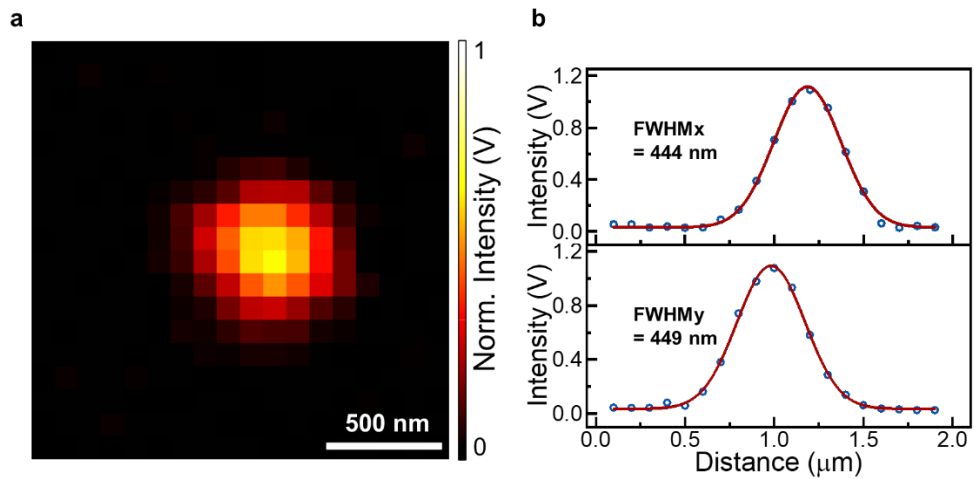

**Fig. S7. Characterization of the SEPT's spatial resolution.** (A) Single AuNP imaged by SEPT microscope. (B) The line profile and the Gaussian fitting results.

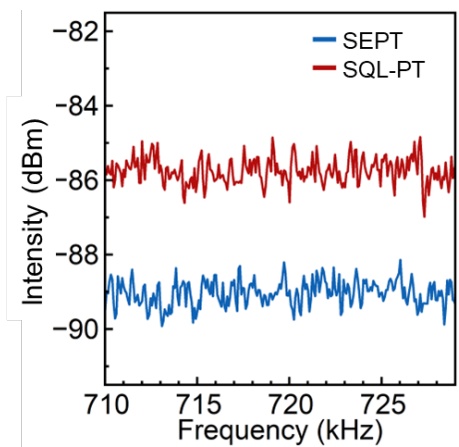

**Fig. S8. Signal baseline of SEPT and SQL-PT without pump heating.** The RBW is 300 Hz, VBW is 30 Hz. Each spectrum is measured with 5 s and averaged of 10 times.

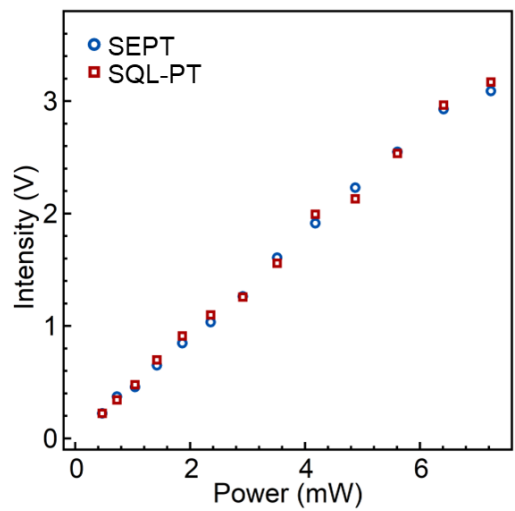

**Fig. S9. Pump-power dependence of photothermal signal amplitudes in SEPT and SQL-PT.**

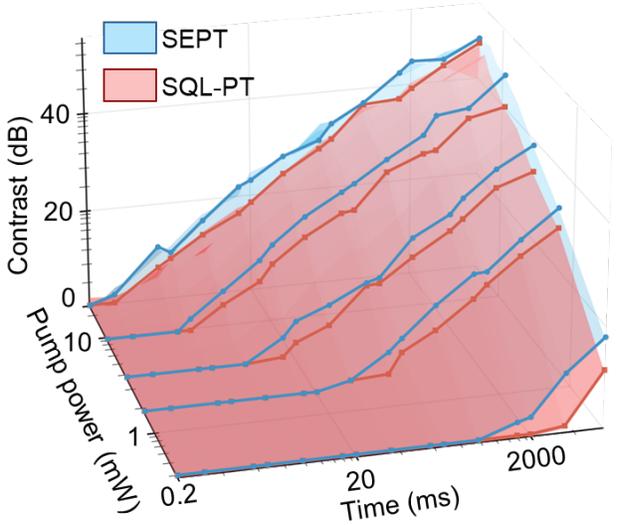

**Fig. S10. Comparison of image contrast between SEPT and SQL-PT**

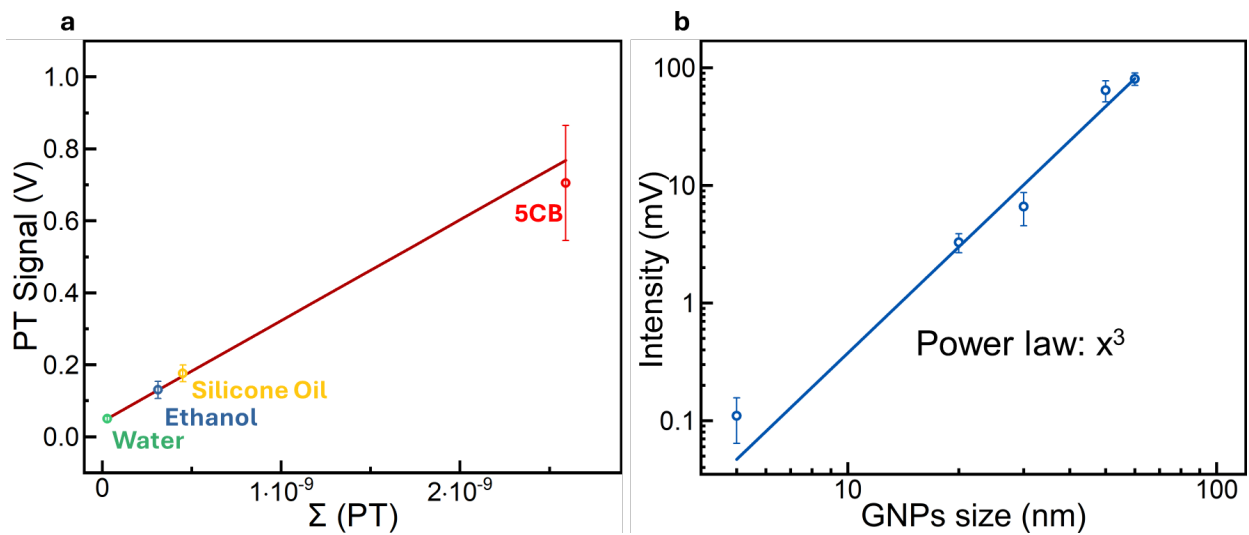

**Fig. S11. Validation of reliability of our PT system.** (**A**) PT signal of AuNPs at different embedding medium. (**B**) Signal size dependence of AuNPs at silicone oil. It exhibits a good qualitative agreement with a third-order law of the radius of AuNPs (absorption cross section).

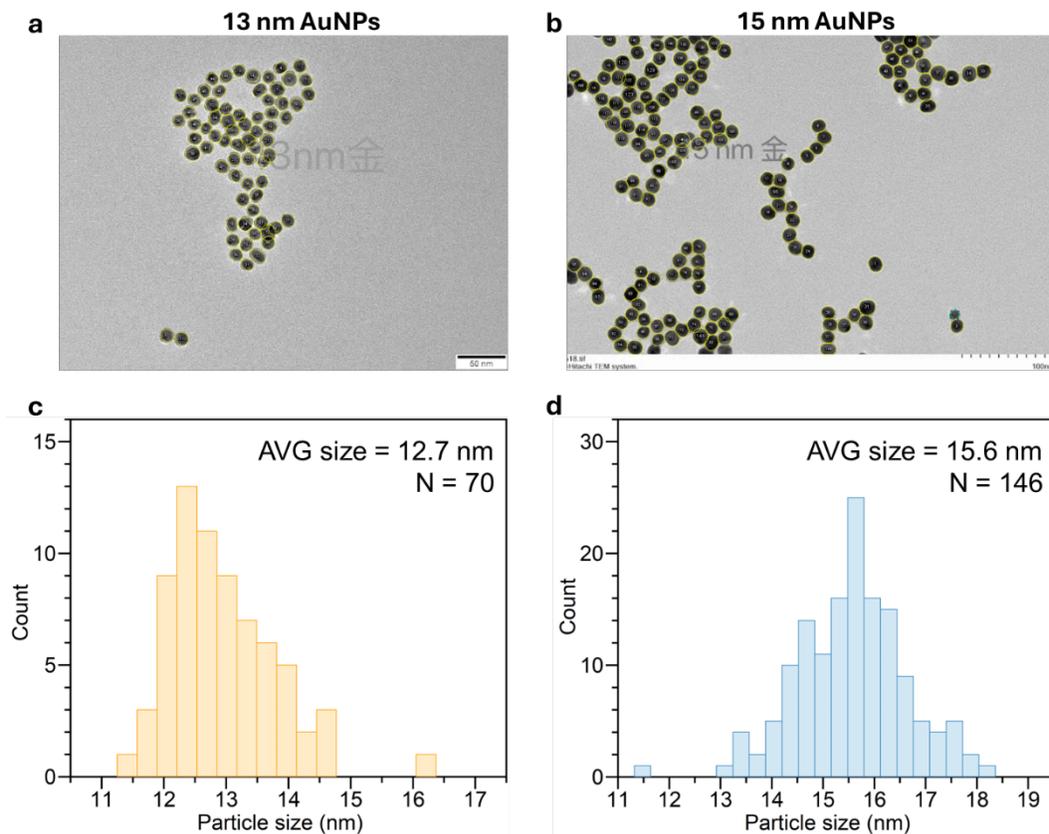

**Fig. S12. Size analysis of AuNPs.** (A and B) Transmission electron microscopy images of AuNPs with labeled sizes (~13 nm and ~15 nm). (C and D) Size distribution histograms of the 13-nm and 15-nm AuNP clusters, showing average diameters of 12.7 nm and 15.6 nm (measured by fitting of Normal distribution), respectively.

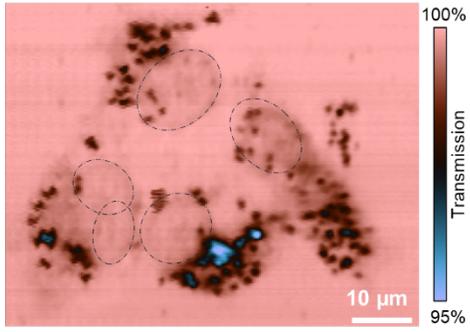

**Fig. S13. Transmission imaging of HEK cells.**

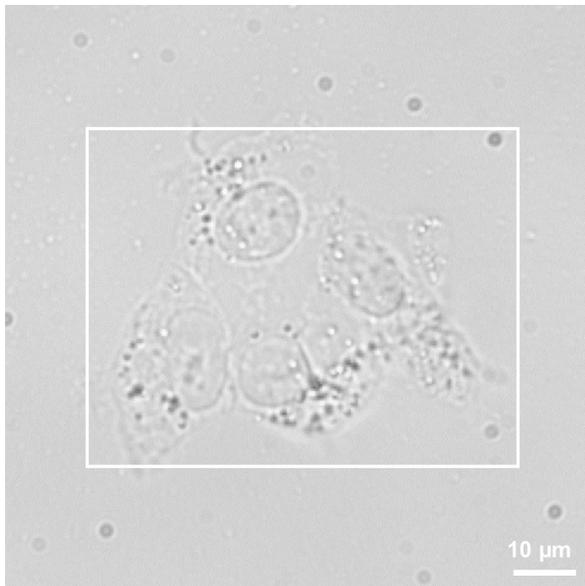

**Fig. S14. Brightfield imaging of HEK cells with imaging field of view indicated by the white box.**

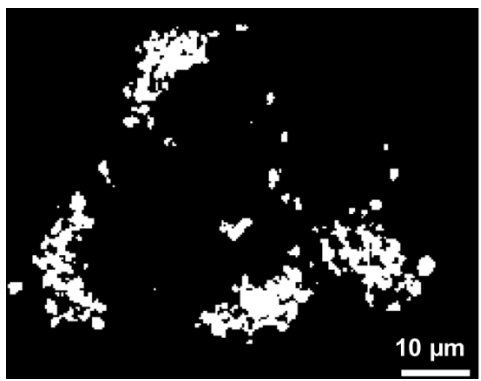

**Fig. S15. Mask for signal extraction for both SEPT and SQL-PT images.**

**References**
1. Gaiduk, A., Ruijgrok, P. V., Yorulmaz, M. & Orrit, M. Detection limits in photothermal microscopy. *Chem. Sci.* **1**, 343-350 (2010).
2. Berciaud, S., Lasne, D., Blab, G. A., Cognet, L. & Lounis, B. Photothermal heterodyne imaging of individual metallic nanoparticles: Theory versus experiment. *Phys. Rev. B* **73**, 045424 (2006).
3. Gaiduk, A., Yorulmaz, M., Ruijgrok, P. & Orrit, M. Room-temperature detection of a single molecule's absorption by photothermal contrast. *Science* **330**, 353-356 (2010).
4. West, C. A. *et al.* Nonlinear effects in single-particle photothermal imaging. *J. Chem. Phys.* **158**, 024202 (2023).
5. Radunz, R., Rings, D., Kroy, K. & Cichos, F. Hot Brownian particles and photothermal correlation spectroscopy. *J. Phys. Chem. A* **113**, 1674-1677 (2009).
6. Rings, D., Schachoff, R., Selmke, M., Cichos, F. & Kroy, K. Hot brownian motion. *Phys. Rev. Lett.* **105**, 090604 (2010).
7. Zharov, V. P. Ultrasharp nonlinear photothermal and photoacoustic resonances and holes beyond the spectral limit. *Nat. Photonics* **5**, 110-116 (2011).
8. Nedosekin, D. A., Galanzha, E. I., Dervishi, E., Biris, A. S. & Zharov, V. P. Super‐resolution nonlinear photothermal microscopy. *Small* **10**, 135-142 (2014).
9. McCormick, C., Marino, A. M., Boyer, V. & Lett, P. D. Strong low-frequency quantum correlations from a four-wave-mixing amplifier. *Physical Review A—Atomic, Molecular, and Optical Physics* **78**, 043816 (2008).
10. Qin, Z. *et al.* Experimental generation of multiple quantum correlated beams from hot rubidium vapor. *Phys. Rev. Lett.* **113**, 023602 (2014).